\documentclass[sigconf, authorversion=true, nonacm=true]{acmart}

\AtBeginDocument{%
  }
\newcommand{\matr}[1]{\mathbf{#1}} 
\RequirePackage{tikz}
\definecolor{orange}{RGB}{233,113,50}
\definecolor{light-blue}{RGB}{0,176,240}
\newcommand{\tikzcircle}[2][red,fill=red]{\tikz[baseline=-0.5ex]\draw[#1,radius=#2] (0,0) circle ;}
\usepackage{multirow}
\usepackage{subcaption}
\usepackage{balance}

\begin{document}

\title[Weak-Annotation of HAR Datasets using Vision Foundation Models]{Weak-Annotation of HAR Datasets using \\Vision Foundation Models}

\author{Marius Bock}
\orcid{0000-0001-7401-928X}
\affiliation{%
  \department{Ubiquitous Computing \&}
  \department{Computer Vision}
  \institution{University of Siegen}
  \city{Siegen}
  \country{Germany}
}
\email{marius.bock@uni-siegen.de}

\author{Kristof Van Laerhoven}
\orcid{0000-0001-5296-5347}
\affiliation{%
  \department{Ubiquitous Computing}
  \institution{University of Siegen}
  \city{Siegen}
  \country{Germany}
}
\email{kvl@eti.uni-siegen.de}

\author{Michael Moeller}
\orcid{0000-0002-0492-6527}
\affiliation{%
  \department{Computer Vision}
  \institution{University of Siegen}
  \city{Siegen}
  \country{Germany}
}
\email{michael.moeller@uni-siegen.de}

\begin{abstract}
As wearable-based data annotation remains, to date, a tedious, time-consuming task requiring researchers to dedicate substantial time, benchmark datasets within the field of Human Activity Recognition in lack richness and size  compared to datasets available within related fields. Recently, vision foundation models such as CLIP have gained significant attention, helping the vision community advance in finding robust, generalizable feature representations. With the majority of researchers within the wearable community relying on vision modalities to overcome the limited expressiveness of wearable data and accurately label their to-be-released benchmark datasets offline, we propose a novel, clustering-based annotation pipeline to significantly reduce the amount of data that needs to be annotated by a human annotator. We show that using our approach, the annotation of centroid clips suffices to achieve average labelling accuracies close to 90\% across three publicly available HAR benchmark datasets. Using the weakly annotated datasets, we further demonstrate that we can match the accuracy scores of fully-supervised deep learning classifiers across all three benchmark datasets. Code as well as supplementary figures and results are publicly downloadable via \url{github.com/mariusbock/weak_har}.
\end{abstract}

\ccsdesc[500]{Human-centered computing~Ubiquitous and mobile computing design and evaluation methods}

\keywords{Data Annotation; Human Activity Recognition; Body-worn Sensors}

\begin{teaserfigure}
    \centering
	\includegraphics[width=0.95\textwidth]{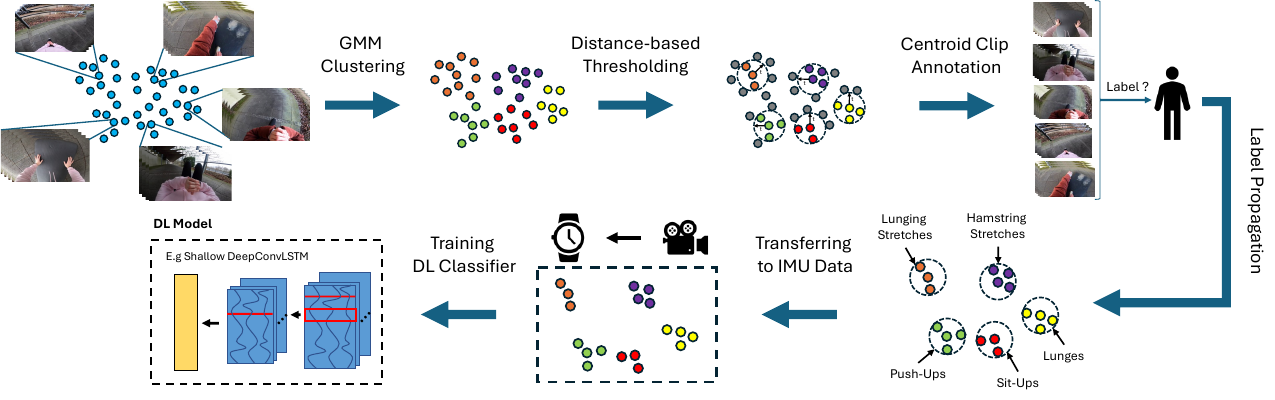}
    \caption{Our proposed weak-annotation pipeline: Visual embeddings extracted using Vision Foundation Models are clusterd using Gaussian Mixture Models (GMMs). Decreasing the required labelling effort, a human annotator is only asked to annotate each cluster's centroid video clip. Centroid labels are then propagated within each cluster. Transferred to the corresponding IMU-data, resulting weakly-annotated datasets can be used to train subsequent classifiers.}
    \label{fig:banner}
    \Description{Visualization of our proposed weak-annotation pipeline: Visual embeddings extracted using Vision Foundation Models are clusterd using Gaussian Mixture Models (GMMs). Decreasing the required labelling effort, a human annotator is only asked to annotate each cluster's centroid video clip. Centroid labels are then propagated within each cluster. Transferred to the corresponding IMU-data, resulting weakly-annotated datasets can be used to train subsequent classifiers.}
\end{teaserfigure}

\maketitle

\section{Introduction}
Though the automatic recognition of activities through wearable data has been identified as valuable information for numerous research fields \cite{bullingTutorialHumanActivity2014}, currently available wearable activity recognition benchmark datasets lack richness and size compared to datasets available within related fields. Compared with, for example, the newly released Ego4D dataset \cite{graumanEgo4DWorld0002022}, it becomes apparent that currently used datasets within the inertial-based Human Activity Recognition (HAR) community are significantly smaller in terms of the number of participants, length of recordings, and variety of performed activities. One of the main drivers for this is that, even though body-worn sensing approaches allow recording large amounts of data with only minimal impact on users in various situations in daily life, wearable-based data annotation remains, to date, a tedious, time-consuming task and requires researchers to dedicate substantial time to it during data collection (taking up to 14 to 20 times longer than the actual recorded data \cite{roggenCollectingComplexActivity2010}).

Following the success in other vision-related fields, researchers within the video activity recognition community have made use of feature extraction methods which provide latent representations of video clips rather than using raw image data \cite{shiTriDetTemporalAction2023, liuEndToEndTemporalAction2022}. Such feature extraction methods are usually pretrained on existing large benchmark corpora, though often not particularly related, they are capable of transferring knowledge to the activity recognition task at hand. Recently, vision foundation models \cite{radfordLearningTransferableVisual2021, oquabDINOv2LearningRobust2024} have gained a lot of attention. Typically trained on a large amount of curated and uncurated benchmark datasets, these models have helped the community further advance in finding robust, generalizable visual feature representations.

With the majority of researchers within the wearable activity recognition community relying on the vision modality to overcome the lack expressiveness of wearable data and accurately label their to-be-released benchmark datasets offline (see e.g. \cite{chanCAPTURE24LargeDataset2024, roggenCollectingComplexActivity2010, hoelzemannHangTimeHARBenchmark2023}), we propose a novel annotation pipeline which makes use of visual embeddings extracted using pretrained foundation models to significantly limit the amount of data which needs to be annotated by a human annotator. Our contributions are three-fold:

\begin{enumerate}
    \item We find that visual embeddings extracted using publicly-available vision foundation models
    can be clustered activity-wise.
    \item We show that annotating only one clip per cluster suffices to achieve average labelling accuracies above 60\% and close to 90\% across three publicly available HAR benchmark datasets. 
    \item We demonstrate that using the weakly annotated datasets, one is capable of matching accuracy scores of fully-supervised deep learning classifiers across all three benchmark datasets.
\end{enumerate}

\section{Related Work}

\paragraph{Vision Foundation Models}
The term foundation models was coined by \citet{devlinBERTPretrainingDeep2019} and refers to models which are pre-trained on a large selection of datasets. The idea of pre-training models on large benchmark datasets has been prominent within the vision community for a long time. Within the video classification community researchers demonstrated that pretrained methods such as I3D \cite{carreiraQuoVadisAction2017}, VideoSwin \cite{liuVideoSwinTransformer2022} or SlowFast \cite{feichtenhoferSlowFastNetworksVideo2019} extract discriminate feature embeddings which can be use to train subsequent classifiers. Following their success in Natural Language Processing \cite{brownLanguageModelsAre2020, devlinBERTPretrainingDeep2019}, researchers applied masked autoencoders on visual data input. Unlike previous methods, masked autoencoders are capable of pretraining themselves in a self-supervised manner, allowing the use of larger data sources. Two of such methods are CLIP \cite{radfordLearningTransferableVisual2021} and DINOv2 \cite{oquabDINOv2LearningRobust2024}. The former, published by OpenAI, is a vision-language model which tries to learn the alignment between text and images. According to the authors, CLIP is pretrained on a large copora of image-text pairs scraped from the world wide web. Similarly, the recently released DINOv2 by META AI makes an effort of providing a foundation model which is capable of extracting general purpose visual features, for which the authors collected data from curated and uncurated sources.

\paragraph{Weakly-Supervised Wearable HAR}

With the activity labelling of body-worn sensor data being a tedious task, many researchers have looked at weakly-supervised learning techniques to reduce the required amount of annotations to train subsequent classifiers. Early works such that of \citet{stikicWeaklySupervisedRecognition2011} have shown to reduce the labelling efforts for training classical machine learning models through knowledge-driven approaches using graph-based label propagation \cite{stikicMultigraphBasedSemisupervised2009}, multi-instance learning \cite{stikicActivityRecognitionSparsely2009} or probabilistic methods \cite{ghazvininejadHMMBasedSemisupervised2011, tanhaSemisupervisedSelftrainingDecision2017}. \citet{adaimiLeveragingActiveLearning2019} followed the works of \cite{stikicWeaklySupervisedRecognition2011}, proposing an active learning framework which focuses on asking users to label only data which will gain the most classification performance boost. With the rise in popularity of Deep Learning, deep clustering algorithms have been proposed to cluster latent representations in unsupervised and semi-supervised fashion using e.g. autoencoders \cite{liUnsupervisedFeatureLearning2014, wangHumanActivityRecognition2016, almaslukhEffectiveDeepAutoencoder2017, maUnsupervisedHumanActivity2021, azadiRobustFeatureRepresentation2024}, recurrent networks \cite{ghazvininejadHMMBasedSemisupervised2011, abedinDeepClusteringHuman2020}, self-supervised \cite{saeedMultitaskSelfSupervisedLearning2019} and contrastive learning \cite{ahmedClusteringHumanActivities2022, xiaTS2ACTFewShotHuman2023, tongZeroShotLearningIMUBased2021, deldariCOCOACrossModality2022, khaertdinovContrastiveSelfsupervisedLearning2021}. Recently, \citet{xiaTS2ACTFewShotHuman2023} and \citet{tongZeroShotLearningIMUBased2021} demonstrated how vision-foundation models such as CLIP \cite{radfordLearningTransferableVisual2021} and I3D \cite{carreiraQuoVadisAction2017} can be used to create visual, complementary embeddings to inertial data such that a contrastive loss can be calculated. This work marks one of the few instances of researchers trying to use visual data to limit the amount of annotations required in wearable activity recognition. Our work ties into the works of \citet{tongZeroShotLearningIMUBased2021} and \citet{xiaTS2ACTFewShotHuman2023}, yet we propose instead to apply vision foundation models to perform automatic label propagation between similar embeddings.

\section{Methodology}

\subsection{Annotation Pipeline}

\paragraph{Latent Space Clustering via Vision Foundation Models}

Within the first phase we divide the unlabeled dataset into (overlapping) video clips. Given an input video stream $X$ of a sample participant, we apply a sliding window approach which shifts over $X$, dividing the input data into video clips, e.g. of four second duration with a 75\% overlap between consecutive windows. This process results in $X = \{\matr{x}_1, \matr{x}_2, ..., \matr{x}_T\}$ being discretized into $t = \{0, 1, ..., T\}$ time steps, where $T$ is the number of windows, i.e. video clips, for each participant. Inspired by classification approaches originating from the temporal action localization community, we make use of pretrained vision foundation models to extract latent representations of each clip. That is, $\mathbf{x}_t\in \mathbb{R}^{E}$ represents a one-dimensional feature embedding vector associated with the video clip at time step $t$, where $E$ is number of latent features the embedding vector consists of. In total we evaluated three popular pretrained foundation models: a two-stream inflated 3D-ConvNet (I3D) \cite{carreiraQuoVadisAction2017} pretrained on the RGB and optical flow features extracted from Kinetics-400 dataset \cite{kayKineticsHumanAction2017} as well as two transformer foundation models CLIP \cite{radfordLearningTransferableVisual2021} and DINOv2 \cite{oquabDINOv2LearningRobust2024}, which were pretrained on a multitude of curated and uncurated data sources. Note that, unlike Carreira and Zisserman in \cite{carreiraQuoVadisAction2017}, we use RAFT \cite{teedRAFTRecurrentAllPairs2020} instead of TV-L1 \cite{sanchezperezTVL1OpticalFlow2013} optical flow estimation. As the CLIP and DINOv2 model both are not explicitly trained on optical flow features, we also test complementing embeddings of the two models by concatenating them with extracted embeddings of the inflated 3D-ConvNet trained on RAFT optical flow features of the Kinetics dataset. In order to obtain latent representations we altered models such that intermediate feature representations can be extracted. Table \ref{tab:extraction} provides details activations at which layer were considered to be the embedding of which pretrained method as well as their dimension. To merge together the frame-wise features outputted of the CLIP and DINOv2 model, we apply average pooling as detailled in \cite{luoCLIP4ClipEmpiricalStudy2022} to obtain a single latent representation per sliding video clip.

\begin{table}
  \caption{Network layer used for extracting embeddings of the different vision foundation models \cite{carreiraQuoVadisAction2017, radfordLearningTransferableVisual2021, oquabDINOv2LearningRobust2024}. Subsequent layers are omitted such that the network outputs latent representations at point of the embedding layer. Note that the I3D network is used for extracting both RGB and flow features and we refer to the vision-based part of the CLIP model.}
  \label{tab:extraction}
  \begin{tabular}{llc}
    \toprule
    Model      & Embedding Layer & Dimension \\
    \midrule
    I3D & last average pool layer & $\mathbb{R}^{1024}$ \\
    CLIP       & last projection layer (vision-CLIP) & $\mathbb{R}^{768}$ \\
    DINO     & last layer hidden state (clf. token) & $\mathbb{R}^{1024}$ \\
  \bottomrule
\end{tabular}
\end{table}

Having extracted latent representations of each video clip within the tested benchmark datasets, we apply Gaussian Mixture Models (GMM) to cluster the embeddings on a per-participant level. Though GMMs are not originally intended to be used with high-dimensional data, they have shown to provide good results clustering visual embeddings especially in the context of action recognition \cite{kuklevaUnsupervisedLearningAction2019, vidalmataJointVisualTemporalEmbedding2021} and allow, unlike methods such as k-nearest neighbors, more flexibility regarding the shape of clusters. Training one GMM clustering algorithm per participant and applying it to said that assigns each video clip  $\mathbf{x}_t \in \mathbb{R}^{E}$ a cluster label $x_c \in {1, ..., C}$, where $C$ is the number of GMM components, i.e. clusters, applied.

\paragraph{Weak-Labeling via Centroid Clips}

Once each video clip of a study participant has been assigned a cluster label $x_c$, the second phase of our approach consists of a human annotator only needing to annotate one sample clip per cluster. Assuming the centroid of a cluster is most representative of all clips within that cluster, we can propagate the activity label $a \in {1, ..., A}$ of said clip to all other clips eliminating the need of annotating the other clips via a human annotator, where $A$ is the number of activities within the dataset. As GMM do not explicitly provide a definition of a centroid of a component, we calculate the centroid clip of each cluster component being the clip which has the highest density within said cluster. That is, given the covariance matrix of each mixture component $\Sigma \in \mathbb{R}^{E \times E}$, assuming each component has its own general covariance matrix, and mean vector $\mu \in \mathbb{R}^{E}$, we calculate the density of each point as the logarithm of the probability density function of the multivariate normal distribution defined by $\mu$ and $\Sigma$. Having identified the centroid clip within each cluster, our approach propagates the annotation provided by the human annotator to all other clips, which were also assigned to that cluster. 

As our approach forces each video clip to be assigned an activity label, we augment our clustering with a subsequent distance-based thresholding in order to remove outlier clips from the automatic labelling. Assuming that the distance of another clip to the centroid resembles its likelihood of belonging to the same activity class, we omit clips from the dataset which exceed a certain distance from their respective centroid clip, with the distance being calculated as the $L^2$-norm between two embedding vectors. Even though this approach decreases the amount of data which can be used to train subsequent classification algorithms, we show to increase the overall labelling accuracy by a significant margin.

\begin{figure*}
  \centering
  \includegraphics[width=\linewidth]{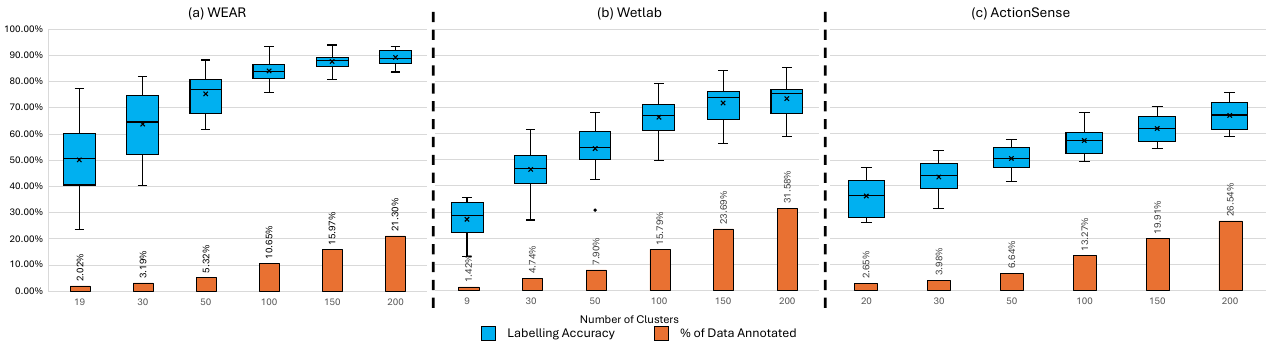}
  \caption{Box-plot diagrams \tikzcircle[fill=light-blue]{2.3pt} showing the distribution of labelling accuracies  across study participants with increasing number of clusters. The bar plot \tikzcircle[fill=orange]{2.3pt} below the box-plots provides details per cluster setting about the percentage of data compared to the total size of the three benchmark datasets \cite{bockWEAROutdoorSports2023, schollWearablesWetLab2015, delpretoActionSenseMultimodalDataset2022} an annotater would need to annotate. One can see a clear trend that with an increase in clusters, labelling accuracy increases along with deviation across study participants decreasing.}
  \label{fig:noclusters}
  \Description{Three box-plot and bar-plot diagrams comparing the average labelling accuracy and standard deviation across participants applying a different amount of clusters.}
\end{figure*}

\subsection{Weakly-supervised Training}

Assuming inertial and video data are synchronised, we further evaluate how well the resulting annotated inertial data with non-uniform label noise is suited to be used for training inertial-based deep learning classifiers. As our benchmark algorithms of choice we use two recently published state-of-the-art methods, namely the Shallow DeepConvLSTM \cite{bockImprovingDeepLearning2021} and TinyHAR architecture \cite{zhouTinyHARLightweightDeep2022}. We use both architectures as originally introduced by the authors, specifically using the same size and number of convolutional filters, convolutional layers and size of the recurrent layers. During training we apply a sliding window of one second with an overlap of 50\%, as it proofed to be provide consistent classification performances across a multitude of HAR datasets \cite{bockImprovingDeepLearning2021}. We train each network for 30 epochs using the Adam optimizer (learning rate $1e^{-4}$ and weight decay $1e^{-6}$) applying a step-wise learning rate with a decay factor of 0.9 after each 10 epochs. To migitate the introduced label noise by our proposed weak-annotation pipeline, we calculate the loss during training using the weighted partially Huberised generalised cross-entropy (PHGCE) loss \cite{menonCanGradientClipping2020}, which extends the definition of the generalized cross-entropy loss \cite{zhangGeneralizedCrossEntropy2018} with a variant of gradient clipping. 
To compare the validity of our approach, we compare amongst a set of (weakly-)annotated training approaches:
\begin{enumerate}
    \item \textit{Fully-supervised:} Fully-supervised results using the original, fully-annotated benchmark datasets.
    \item \textit{Few-Shot-CE:} Fully-supervised training using only the annotated clips and a weighted cross-entropy loss.
    \item \textit{Random-CE}:  training using an equal amount of random annotated clips as in (2) and a weighted cross-entropy loss.
    \item \textit{Weak-CE:} Weakly-supervised training using the weakly-annotated dataset and a weighted cross-entropy loss.
    \item \textit{Weak-PHGCE:} Weakly-supervised training using the weakly-annotated dataset and a weighted PHGCE loss.
\end{enumerate}

\begin{table}[!htb]
  \caption{Average labeling accuracy and standard deviation across study participants using different types and combinations of embeddings \cite{carreiraQuoVadisAction2017, radfordLearningTransferableVisual2021, oquabDINOv2LearningRobust2024} extracted from three benchmark datasets \cite{bockWEAROutdoorSports2023, schollWearablesWetLab2015, delpretoActionSenseMultimodalDataset2022} applying a GMM-based clustering using 100 clusters. Overall, a combination of CLIP and optical flow embeddings proved most consistent across all datasets.}
  \label{tab:embeddings}
  \begin{tabular}{lccc}
    \toprule
                        & WEAR    & Wetlab & ActionSense\\
    \midrule
    (1) I3D            & 82.62 ($\pm 4.65$)          & 66.08 ($\pm$ 9.53)          & 53.47 ($\pm$ 5.95) \\
    (2) CLIP           & 82.47 ($\pm 6.03$)          & \textbf{72.70 ($\pm$ 6.42)} & 59.85 ($\pm$ 4.42) \\
    (3) DINOv2         & 79.20 ($\pm 4.04$)          & 69.28 ($\pm$ 8.12)          & \textbf{60.25 ($\pm$ 4.04)} \\   
    (4) RAFT           & 76.86 ($\pm 4.79$)          & 51.50 ($\pm$ 6.96)          & 45.64 ($\pm$ 5.19) \\
    (1) + (4)          & \textbf{85.17 ($\pm$ 4.48)} & 60.91 ($\pm$ 8.36)          & 53.00 ($\pm$ 4.66) \\
    (2) + (4)          & 83.96 ($\pm 4.99$)          & 66.23 ($\pm 7.86$)          & 57.29 ($\pm$ 5.64) \\
    (3) + (4)          & 79.13 ($\pm 4.30$)          & 70.18 ($\pm$ 9.79)          & 56.55 ($\pm$ 4.51) \\   
  \bottomrule
\end{tabular}
\end{table}

\subsection{Datasets}

\paragraph{WEAR} The WEAR dataset offers both inertial and egocentric video data of 18 participants doing a variety of 18 sports-related activities, including different styles of running, stretching, and strength-based exercises. Recordings took place at changing outdoor locations. Each study participant was equipped with a head-mounted camera and four smartwatches, one worn on each limb in a fixed orientation, which captured 3D-accelerometer data.

\paragraph{ActionSense} Published by \citet{delpretoActionSenseMultimodalDataset2022}, the ActionSense dataset provides a multitude of sensors capturing data within an indoor, artificial kitchen setup. Amongst the sensors, participants wore Inertial Measurment Units (IMUs) on both wrists as well as smart glasses which captured the ego-view of each participant. During recordings, participants were tasked to perform various kitchen chores including chopping foods, setting a table and (un)loading a dishwasher. Within their original publication, the authors provide annotations of 19 activities of 10 participants. Note that the dataset download of the ActionSense dataset provides IMU and egocentric video data of only 9 instead of 10 participants.

\paragraph{Wetlab} Taking place in a wetlab laboratory environment, the Wetlab dataset \cite{schollWearablesWetLab2015} comprises of data of 22 study participants which performed two DNA extraction experiments. For purposes of this paper we used the annotated provided by the authors of the reoccurring activities base activities (such as stirring, cutting, etc.) within the experimental protocol. During recordings, each participant wore a smartwatch in a fixed orientation on the wrist of their dominant hand, which captured 3D-accelerometer data. Unlike the WEAR and ActionSense dataset, the Wetlab dataset  provides video data of a static camera which was mounted above the table at which the experiment was performed, thus capturing a birds-eye perspective of the experiment's surroundings.

\section{Results}
To ensure that reported performance differences are not the based on statistical variance, all reported experiments are repeated three times, applying a set of three predefined random seeds. This applies both for the annotation pipeline experiments as well as weakly-supervised training results. During all annotation-based experiment mentioned in Section \ref{sec:annoresults} we apply a clip length of four seconds along with a three second overlap between clips. We assume that this clip length is a suitable length to be interpretable for a human annotator while simultaneously avoiding mixing multiple activities into one sliding window. Furthermore, during all experiments only the label of the centroid clip is propagated to that of all other cluster instances. Ablation experiments evaluating different clip lengths and number of annotated clips per cluster to determine the label to be propagated can be found within our code repository.

\begin{table*}
  \caption{Deep Learning results of applying two inertial-based models \cite{bockImprovingDeepLearning2021, zhouTinyHARLightweightDeep2022} on various weakly-annotated versions of three public datasets \cite{bockWEAROutdoorSports2023, schollWearablesWetLab2015, delpretoActionSenseMultimodalDataset2022}. Training using weakly-annotated datasets outperformed both few-shot training using only the annotated data as well as an equal amount of random annotated clips. With an increase in number of clusters our weakly-supervised approach is capable of being close to matching the predictive performance of fully-supervised baselines having manually annotated only a fraction of the actual dataset. The suffix \textit{T-6} (\textit{T-4}) refer to training applying a threshold of 6 (4).}
  \small
  \label{tab:deeplearning}
  \begin{tabular}{clcccccccccccccc}
    \toprule
    \multirow{12}{*}{\rotatebox[origin=c]{90}{WEAR}}& & \multicolumn{6}{c}{DeepConvLSTM} & & \multicolumn{6}{c}{TinyHAR} \\
    & & \multicolumn{2}{c}{$c=19$} & \multicolumn{2}{c}{$c=50$} & \multicolumn{2}{c}{$c=100$} & & \multicolumn{2}{c}{$c=19$} & \multicolumn{2}{c}{$c=50$} & \multicolumn{2}{c}{$c=100$} \\
    & & Acc & F1 & Acc & F1 & Acc & F1 & & Acc & F1 & Acc & F1 & Acc & F1\\ \cmidrule{2-15}
    & \textit{Fully-supervised} & \textit{79.89} & \textit{78.36} & \textit{79.89} & \textit{78.36} & \textit{79.89} & \textit{78.36} &  & \textit{77.83} & \textit{71.89} & \textit{77.83} & \textit{71.89} & \textit{77.83} & \textit{71.89} \\ \cmidrule{2-15}
    & Few-Shot-CE & 37.41 & 24.76 & 59.58 & 46.25 & 65.61 & 53.51 &  & 37.41 & 26.55 & 59.58 & 46.25 & 65.61 &  53.51 \\
    & Random-CE & 45.90 & 31.13 & 59.46 & 46.98 & 65.91 & 53.38 &  & 23.73 & 23.73 & 59.72 & 46.34 & 66.27 & 55.00  \\
    & Weak-CE & 42.55 & 34.09 & 64.17 & 54.59 & 73.38 & 63.23 &  & 49.45 & 38.75 & 66.68 & 54.05 & 71.10 & 59.65 \\
    & Weak-PHGCE & 48.62 & 35.45 & 70.34 & 55.27 & 76.15 & 63.43 &  & 51.46 & 39.23 & 68.53 & 55.40 & 73.37 & 61.68  \\
    & Weak-CE-T-6 & 59.70 & 46.63 & 73.06 & 60.60 & 76.28 & 66.13 &  & 57.22 & 46.71 & 68.19 & 55.56 & 72.03 & 60.31 \\
    & Weak-PHGCE-T-6 & 59.39 & 44.97 & 73.17 & 58.84 & 77.55 & 64.77 &  & 58.78 & 46.27 & 69.47 & 56.29 & 74.05 & 61.67 \\    
    & Weak-CE-T-4 & \textbf{68.86} & \textbf{57.33} & \textbf{74.72} & \textbf{63.93} & \textbf{77.81} & \textbf{68.22} &  & \textbf{65.68} & \textbf{55.64} & 71.31 & 60.16 & 73.93 & 63.42 \\
    & Weak-PHGCE-T-4 & 61.35 & 47.00 & 74.45 & 60.61 & 76.64 & 64.81 &  & 62.90 & 50.46 & \textbf{72.25} & \textbf{60.84} & \textbf{74.83} & \textbf{63.94} \\    
    \midrule
    \multirow{11}{*}{\rotatebox[origin=c]{90}{Wetlab}} & & \multicolumn{2}{c}{$c=9$} & \multicolumn{2}{c}{$c=50$} & \multicolumn{2}{c}{$c=100$} & & \multicolumn{2}{c}{$c=9$} & \multicolumn{2}{c}{$c=50$} & \multicolumn{2}{c}{$c=100$} \\
    & & Acc & F1 & Acc & F1 & Acc & F1 & & Acc & F1 & Acc & F1 & Acc & F1\\ \cmidrule{2-15}
    & \textit{Fully-supervised} & \textit{45.27} & \textit{38.64} & \textit{45.27} & \textit{38.64} & \textit{45.27} & \textit{38.64} &  & \textit{38.75} & \textit{28.85} & \textit{38.75} & \textit{28.85} & \textit{38.75} & \textit{28.85} \\ \cmidrule{2-15}
    & Few-Shot-CE & 15.60 & 11.39 & 21.89 & 16.46 & 22.78 & 17.38 &  & 15.18 & 11.62 & 22.43 & 16.14 & 25.92 & 18.27 \\
    & Random-CE & 16.33 & 8.95 & 26.48 & 18.62 & 26.50 & 20.05 &  & 34.23 & 24.38 & 27.74 & 17.79 & 29.70 & 19.37 \\
    & Weak-CE & 16.97 & 8.53 & 32.57 & 25.78 & 36.51 & 29.72 &  & 23.90 & 14.70 & \textbf{34.23} & \textbf{24.38} & \textbf{36.30} & \textbf{25.76} \\
    & Weak-PHGCE & 18.62 & 10.48 & 27.64 & 23.17 & 33.78 & 27.62 &  & 24.06 & 15.19 & 33.79 & 24.35 & 35.53 & 25.53  \\
    & Weak-CE-T-6 & 23.01 & 18.12 & \textbf{33.78} & \textbf{27.08} & \textbf{38.41} & \textbf{29.77} &  & \textbf{26.14} & \textbf{18.63} & 33.79 & 24.21 & 36.20 & 25.76  \\
    & Weak-PHGCE-T-6 & 20.26 & 13.72 & 28.77 & 23.90 & 32.29 & 26.71 &  & 25.22 & 18.41 & 33.34 & 24.25 & 34.93 & 25.17  \\
    & Weak-CE-T-4 & \textbf{23.54} & \textbf{18.71} & 32.02 & 26.37 & 35.25 & 28.95 &  & 25.15 & 17.91 & 32.85 & 23.84 & 34.64 & 25.04 \\
    & Weak-PHGCE-T-4 & 21.82 & 15.81 & 28.33 & 23.91 & 31.20 & 25.64 &  & 23.82 & 17.29 & 32.39 & 23.72 & 33.66 & 24.70 \\
    \midrule
    \multirow{11}{*}{\rotatebox[origin=c]{90}{ActionSense}}& & \multicolumn{2}{c}{$c=20$} & \multicolumn{2}{c}{$c=50$} & \multicolumn{2}{c}{$c=100$} & & \multicolumn{2}{c}{$c=20$} & \multicolumn{2}{c}{$c=50$} & \multicolumn{2}{c}{$c=100$} \\
    & & Acc & F1 & Acc & F1 & Acc & F1 & & Acc & F1 & Acc & F1 & Acc & F1\\ \cmidrule{2-15}
    & \textit{Fully-supervised} & \textit{20.73} & \textit{14.82} & \textit{20.73} & \textit{14.82} & \textit{20.73} & \textit{14.82} &  & \textit{22.19} & \textit{18.04} & \textit{22.19} & \textit{18.04} & \textit{22.19} & \textit{18.04} \\ \cmidrule{2-15}
    & Few-Shot-CE & 6.47 & 2.58 & 8.14 & 4.92 & 9.96 & 6.08 &  & 6.19 & 2.32 & 8.50 & 4.00 & 11.15 & 6.72 \\
    & Random-CE & 4.69 & 1.63 & 9.42 & 5.09 & 10.02 & 5.77 &  & 5.01 & 1.70 & 7.80 & 4.19 & 11.44 & 6.50 \\
    & Weak-CE & 12.32 & 7.67 & 13.91 & 9.94 & \textbf{17.35} & \textbf{12.20} &  & 13.74 & \textbf{9.75} & \textbf{15.58} & \textbf{11.55} & \textbf{17.16} & \textbf{13.45} \\
    & Weak-PHGCE & 11.65 & 6.34 & 12.34 & 7.43 & 11.62 & 7.02 &  & \textbf{14.35} & 9.43 & 13.88 & 9.25 & 14.52 & 10.09 \\
    & Weak-CE-T-6 & \textbf{14.67} & \textbf{8.81} & \textbf{15.95} & \textbf{10.07} & 15.29 & 10.69 & & 13.35 & 9.32 & 15.70 & 11.22 & 17.53 & 13.15   \\
    & Weak-PHGCE-T-6 & 9.91 & 4.92 & 10.57 & 5.44 & 11.78 & 6.79 & & 12.17 & 7.52 & 13.72 & 8.16 & 14.53 & 9.01   \\
    & Weak-CE-T-4 & 10.40 & 7.15 & 12.31 & 7.10 & 12.08 & 8.80 &  & 10.80 & 8.02 & 13.65 & 9.24 & 15.03 & 11.16  \\
    & Weak-PHGCE-T-4 & 8.34 & 3.86 & 8.92 & 4.35 & 9.91 & 5.56 &  & 9.24 & 5.82 & 11.07 & 6.09 & 12.11 & 7.35  \\
    \bottomrule
\end{tabular}
\end{table*}

\subsection{Annotation Pipeline}
\label{sec:annoresults}
Table \ref{tab:embeddings} shows the average labelling accuracy averaged across participants obtained when applying our proposed annotation pipeline using various types of extracted visual embeddings. One can see that in case of the WEAR \cite{bockWEAROutdoorSports2023} and ActionSense dataset \cite{delpretoActionSenseMultimodalDataset2022} labelling accuracy can be improved by combining both RGB and optical flow features in case of all embeddings. Overall a combination of CLIP and optical flow features proves to be most consistent across our three benchmark datasets of choice, making it thus our embedding of choice for subsequent experiments. Applying a labelling strategy of only annotating the centroid clip of each cluster, Figure \ref{fig:noclusters} presents a box-plot visualization of applying different number of clusters during the clustering of the participant-wise embeddings. One can see that by only annotating 100 clips per study participant, our proposed annotation pipeline is capable of reaching labelling accuracies above 85\% in case of the WEAR and close to 70\% in case of the Wetlab \cite{schollWearablesWetLab2015} and ActionSense dataset. 

Furthermore, as evident by an overall shrinking boxplot with increasing number of clusters, our approach is becoming more stable with the standard deviation across study participants decreasing in case of all three datasets. As Figure \ref{fig:noclusters} shows, applying a clustering of $C = A$, i.e. as many clusters as there are activities in the dataset, results in the clustering not being capable of differentiating the normal and complex variations of activities, different running styles and null-class from all other activities. In general, we witness a trend that by applying a larger amount of clusters than activities present in the dataset, one gives the GMM clustering enough degrees of freedom to differentiate even activities which share similarities, yet slightly differ from each other. 
Lastly, by distance thresholding clusters and excluding instances which are exceeding a certain distance from their respective centroid, helps increasing the labelling accuracy significantly across all datasets. While a threshold of 4 helps increase the labelling accuracy well above 75\% and even up to 93\% in case of the WEAR dataset, the thresholding omits between 50\% and up to 90\% of the datasets.

\subsection{Weakly-Supervised Training}
As a combination of CLIP and optical-flow-based features proved to be most stable across all three datasets, we chose to use said embedding as basis for our weakly-supervised training. Table \ref{tab:deeplearning} provides an overview across the eight evaluated training scenarios. Our proposed weakly-supervised training is not only capable of outperforming the few-shot training using only the annotated centroid clips, but for the case of applying 100 clusters is close to matching accuracy scores of a fully-supervised training across all three benchmark datasets, for both inertial-based architectures. 
Compared to a normal cross-entropy loss, the PHGCE loss provides more stable results in case of higher label noise, e.g., when not applying a distance-based thresholding and/or applying a smaller number of clusters. In general, the distance-based thresholding significantly improved results across all datasets. Although thresholding significantly reduces the amount of training data, the resulting decrease in overall labelling noise, especially for approaches that applied a lower number of clusters, improved classification results. We provide a detailed overview of the influence of threshoding on labelling accuracy and dataset size within the paper's code repository.

\begin{figure}
  \centering
  \includegraphics[width=0.94\columnwidth]{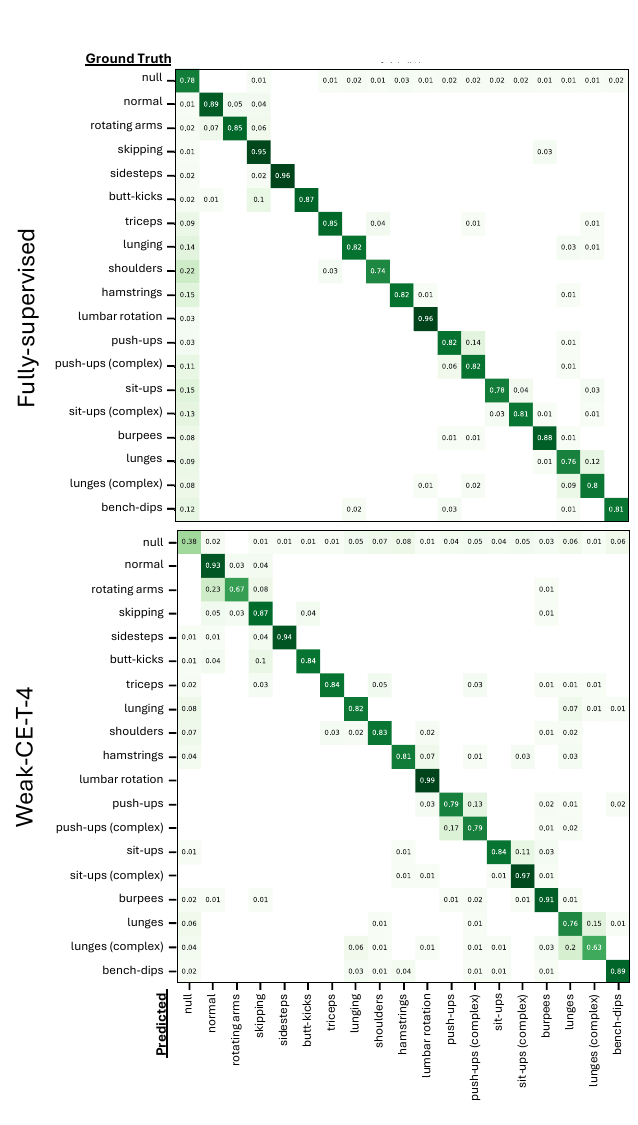}
  \caption{Confusion matrices comparing the shallow DeepConvLSTM fully-supervised results compared to that of the best performing weak-labelling approach. With exception of the NULL-class, all activities were able to be classified close to the performance of the fully-supervised approach.}
  \label{fig:wear_conf}
  \Description{Figure showing 2 confusion matrices comparing applying fully-supervised training with the best weakly-supervised training for the WEAR dataset. One can see that apart from the NULL-Class, the confusion matrices of the fully- and weakly-supervised training share a lot of similarities.}
\end{figure}

Figure \ref{fig:wear_conf} shows a side-by-side comparison of the confusion matrices of the fully-supervised and best-performing weak-labelling approach (Weak-CE-T-4) using the WEAR dataset and shallow DeepConvLSTM. One can see that, apart from the NULL-class, classification results of all activities of the weakly-supervised training are similar to that of the fully-supervised. With the intra-class similarity of a NULL-class within datasets being quite low, it is to a larger degree grouped together with instances of other classes making it harder for an inertial-based classifier to learn unique patterns only applicable to that of the NULL-class. Looking at per-class results of the Wetlab dataset, one can see that the introduced labelling noise caused the classifier trained using weakly-annotated data is more likely to confuse activities with NULL-class rather than vice versa. This caused the activities \textit{transfer} and \textit{pouring}, two classes which only have a few instances in the ground truth data and which are most frequently annotated incorrectly as NULL-class, to not be predicted correctly once across all study participants. Note that for the ActionSense dataset classification results even in the fully-supervised setting are significantly worse compared to that of the other two datasets. As evident by a nevertheless large labelling accuracy using our approach, we assume that label semantics of the dataset are too vision-centric (e.g. peeling a cucumber or a potato) to be correctly recognized using only intertial data. Nevertheless, per-class classification results of the fully- vs. weakly-supervised training show a similar confusion amongst classes, suggesting learned patterns of the classifier are similar in both training scenarios.

\section{Discussion \& Conclusion}
Within this paper we presented a weak-annotation pipeline for HAR datasets based on Vision Foundation Models. We showed that visual embeddings extracted using Vision Foundation Models can be clustered using Gaussian Mixture Models (GMM). Decreasing the required labelling effort, using the suggested pipeline a human annotator is only asked to annotate each cluster's centroid video clip. By propagating the provided labels within each cluster our approach is capable of achieving average labelling accuracies above 60\% and close to 90\% across three popular HAR benchmark datasets. We further showed that the resulting weakly-annotated wearable datasets can be used to train subsequent deep learning classifiers with accuracy scores, in case of applying a sufficiently large number of clusters, being close to matching that of a fully-supervised training across all three benchmark datasets.

Our results underscore one of the implications recent advancements in the vision community in finding generalizable feature representations might have on the field of HAR. With the rapid progress being made in the area of foundation models, follow-ups of models such as CLIP and DINOv2 could further robustify the automatic analysis of collected video streams in wearable-based data collection. Our clustering-based pipeline thus not only improves the efficiency of data annotation but also contributes to the creation of richer and more extensive HAR benchmark datasets.

\begin{acks}
We gratefully acknowledge the DFG Project WASEDO (DFG LA 2758\/11-1) and the University of Siegen's OMNI cluster.
\end{acks}

\bibliographystyle{ACM-Reference-Format}
\balance
\bibliography{main}

\end{document}